\documentclass[twocolumn]{aastex6}
\usepackage{subfig}
\usepackage[utf8]{inputenc}
\usepackage{amsmath}
\usepackage{booktabs}

\newcommand{\teff}{$T_{\textup{eff}}$ }
\newcommand{\vmacro}{v$_{\textup{macro}}$ }
\newcommand{\vsini}{\ensuremath{v \sin{i}} }

\newcommand{\psini}{$P_{\textup{rot}}$/$\sin{i}$ }
\newcommand{\pcrit}{$P_\textup{crit}$ }

\newcommand{\kms}{km~s$^{-1}$ }
\newcommand{\drvmax}{$\Delta RV_{\textup{max}}$ }
\newcommand{\pcritratio}{\frac{P_{\textup{crit}}}{P_{\textup{rot}}/\sin{(i)}} }

\begin{document}
\title{Rotationally Driven Ultraviolet Emission of Red Giant Stars II. Metallicity, Activity, Binarity and Sub-subgiants}
\author{Don Dixon\altaffilmark{1,2}, Keivan G.\ Stassun\altaffilmark{1,2}, Robert D.\ Mathieu\altaffilmark{3}, Jamie Tayar\altaffilmark{4}, Lyra Cao\altaffilmark{1}}

\altaffiltext{1}{Department of Physics and Astronomy, Vanderbilt University, Nashville, TN 37235, USA}
\altaffiltext{2}{Department of Physics, Fisk University, Nashville, TN 37208, USA}
\altaffiltext{3}{Department of Astronomy, University of Wisconsin-Madison, Madison, WI 53706, USA}
\altaffiltext{4}{Department of Astronomy, University of Florida, Gainesville, FL 32611, USA}

\graphicspath{ {./figures} }

\begin{abstract}
Red Giant Branch (RGB) stars are overwhelmingly observed to rotate very slowly compared to main-sequence stars, but a few percent of them show rapid rotation and high activity, often as a result of tidal synchronizationn or other angular momentum transfer events. In this paper we build upon previous work using a sample of 7,286 RGB stars from APOGEE DR17 with measurable rotation. We derive an updated NUV excess vs \vsini rotation-activity relation that is consistent with our previous published version, but reduces uncertainty through the inclusion of a linear [M/H] correction term. We find that both single stars and binary stars generally follow our rotation-activity relation, but single stars seemingly saturate at \psini $\sim$10 days while binary stars show no sign of saturation, suggesting they are able to carry substantially stronger magnetic fields. Our analysis reveals Sub-subgiant stars (SSGs) to be the most active giant binaries, with rotation synchronized to orbits with periods $\lesssim$ 20 days. Given their unusually high level of activity compared to other short-period synchronized giants we suspect the SSGs are most commonly overactive RS CVn stars. Using estimates of critical rotation we identify a handful giants rotating near break-up and determine tidal spin up to this level of rotation is highly unlikely and instead suggest planetary engulfment or stellar mergers in a fashion generally proposed for FK Comae stars.
\end{abstract}

\maketitle

\section{Introduction}\label{section:Introduction}
As cool main-sequence stars evolve they gradually spin down due to angular momentum being carried away by magnetized winds \citep{Weber67}. The boundary (\teff $<$ 6500) where this spin-down mechanism becomes significant separates the vastly different angular momentum evolution between early-type stars with radiative envelopes and late-type stars with convective envelopes \citep{Kraft67,Beyer24}. During the first ascent of the Red Giant Branch (RGB) rotation is slowed even further, a consequence of angular momentum conservation as stars swell in size. Application of these spin-down mechanisms predict that virtually all low-mass evolved stars should rotate very slowly compared to typical main sequence stars. Despite evolutionary spin down, investigation of the field has shown $\sim$2$\%$ of evolved stars exhibit rapid rotation (\vsini $> 10$ \kms) and high levels of activity, which is consistent with population predictions that include spin up via the gravitational influence of a companion star or planet \citep{Ceillier17}. This picture highlights a key difference between fast rotating main-sequence stars and fast rotating giants. For main-sequence stars of a given spectral type rotation and activity connects to age \citep{Skumanich72}, where the practice of deriving ages of cool main-sequence stars from rotation is known as gyrochronology \citep{Barnes03,Barnes07}. However, since giants likely draw the bulk of any significant rotation from the orbital angular momentum of a nearby companion, in their case rotation and activity connects to binary evolution.

The observational properties of a magnetically active giant with rapid rotation induced via the tidal synchronization of a companion is well demonstrated by RS Canum Venaticorum (RS CVn) variable stars. RS CVn stars are short-period binaries where rapid rotation has caused the primary to become chromospherically active, often with starspot variability and/or strong Ca II H \& K emission \citep{Hall76}, and sometimes solar-like activity cycles \citep{Buccino09}. Recently, \citet{Leiner22} demonstrated substantial overlap between stars redder/fainter than the subgiant/giant branch, also known as sub-subgiants (SSGs), and RS CVn systems. Though the formation mechanism for SSGs and how they reach their position in color-magnitude diagrams (CMDs) is still under discussion, the overlap suggests SSGs may be another photometrically observable phase of active giant binary evolution.

In our previous work we use the GALEX/2MASS $NUV - J$ color displacement from an empirically derived color-color locus derived in \citet{Findeisen10} to define an NUV excess parameter as a proxy for stellar activity \citep{Dixon20}. Our analysis of NUV excess found APOGEE-based log($v \sin{i}$) to be strongly linearly correlated with this NUV excess. We also found tentative evidence of saturation/supersaturation regimes of activity using the NUV excess metric. Additionally, we discovered similarity to M-dwarf chromospheric activity when comparing to the sample described in \cite{Stelzer16}. In this work we expand upon our previous paper to deliver a more robust metallicity-calibrated activity metric and additional insight into the role of binary evolution for rapidly rotating giants.  We also find in our sample that SSGs are especially active synchronized giants and that giants near break-up rotation likely formed as a result of swallowing a planet or stellar companion.

The structure of the paper is as follows. In section \ref{section:Sample} we detail the construction of our sample of 7,286 giants used in our study. Section \ref{section:Method} provides an overview of definitions used throughout the paper to provide clarity. In section \ref{section:Results} we describe the details of our analyses and results toward understanding giant rapid rotation and activity. Section \ref{section:Discussion} discusses NUV excess in relation to potential white dwarf companions and directions for future work. Lastly, section \ref{sections:Conclusion} summarizes our major findings.

\section{Sample Construction}\label{section:Sample}
\subsection{APOGEE Spectra and GALEX/Gaia Crossmatch}
In this paper the foundational catalog for sample construction is the primary data table (allstar table) for combined spectra in the 17th data release of the APOGEE sky survey \citep{Abdurro'uf22}. This data release provides high-resolution ($R\sim 22,500)$ spectral observations in the H-band of just over $6.5\times 10^5$ unique systems and has sky coverage of both the northern and southern hemispheres. The allstar table reports stellar parameters for each of these systems, derived from the default APOGEE analysis pipeline ASPCAP \citep{Jonsson20}. 

To start our selection we query for all stars with giant-like surface gravities ($\log g < 3.5$) and a measured \vsini $>$ 0, reducing the 733,901 entries in the full data release to 23,972. These quantities are calculated by ASPCAP for the combined spectrum, where individual visit spectra are joined by resampling each on a logarithmic-spaced wavelength scale.  It should be noted that our \vsini requirement limits our selection to only include stars where the best fit for the combined spectrum is found in the grids of synthetic ``dwarf'' spectra, rather than the grids of synthetic ``giant'' spectra, due to the designated giant grids not including \vsini as a free parameter. The \teff and log g ranges for the dwarf and giant spectral grids overlap with each other, meaning the primary difference is that broadening is modeled as rotation for the dwarf grid and as macroturbulence for the giant grid. In practice this means that any broadening due to macroturbulence (v$_{\textup{macro}}$) in our sample will be absorbed as rotational broadening, because differences between spectral morphology formed via \vsini and \vmacro broadening kernels are largely degenerate at APOGEE resolution. 

Despite this complication we continue to use ASPCAP \vsini for this study, as \vmacro is anticorrelated with both $\log g$ and \teff and is expected to be $<$ 5 \kms for cool low-mass giants \citep{Thygesen12}. In other words, the broadening contribution from \vmacro is too small to misidentify it as rapid rotation and can only contribute a relatively small fraction of large \vsini values, which should limit its impact on analyses throughout the paper. Still, we acknowledge here that \vsini is in fact \vsini + \vmacro.

\begin{figure}[htbp!]
\centering
\includegraphics[width=\linewidth]{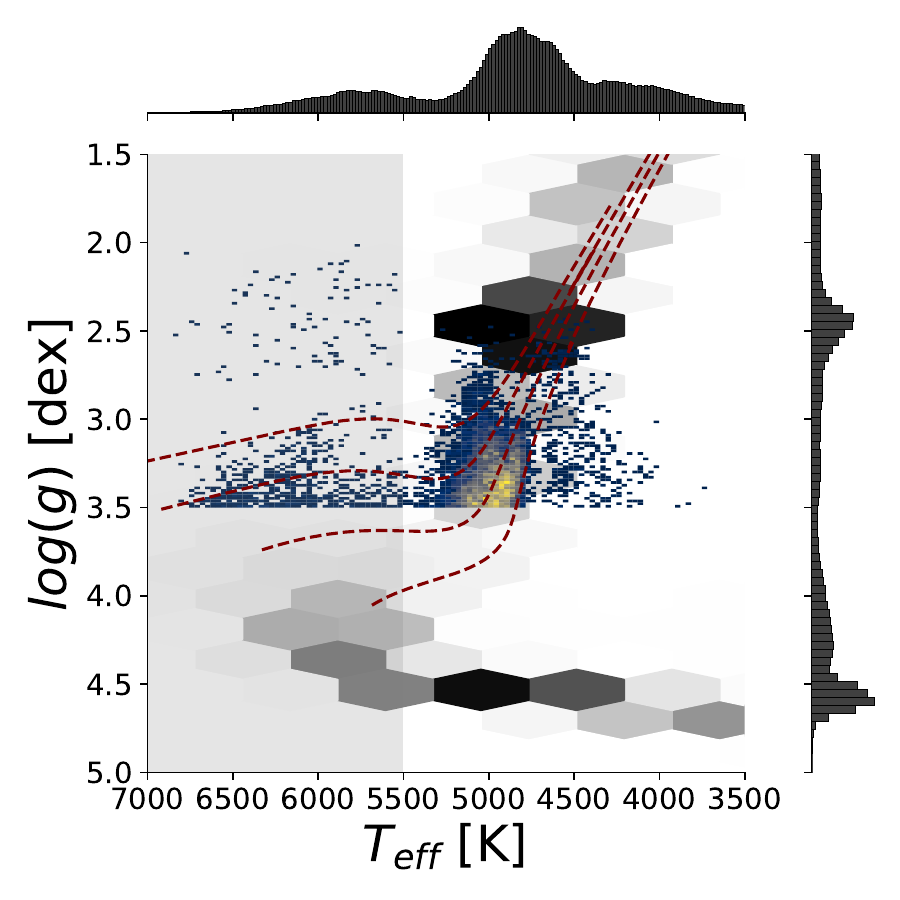}
\caption{Kiel diagram of queried giants from APOGEE DR17. Hexagonal histograms represent the $\sim 6.5\times 10^5$ unique systems in the full catalog. The marginal distributions on the top and right of the figure depict the full sampling of \teff and log g for DR17. Our initial sample of log g $<$ 3.5 dex and \vsini $>$ 0 \kms is plotted as a 2D histogram with 100 bins per axis going from blue to yellow with increasing bin count. The gray background highlights the \teff range where we remove stars not consistent with the giant branch. Partial evolutionary tracks with equal mass spacing are given from 1 M$_\odot$ to 2.5 M$_\odot$.}
\label{fig:kiel_diagram}
\end{figure}

Another reduction in our selected sample results from a \teff criteria to filter stars that lie off the giant branch. Fig. \ref{fig:kiel_diagram} provides an evolutionary illustration of our sample using a spectroscopic HR diagram, also known as a Kiel diagram. In this plot the APOGEE background is represented as an hexagonal 2-dimensional histogram with a histogram of our sample before making the \teff cut overplotted. Additionally, we show MIST stellar evolutionary tracks \citep{Choi16} for 1.0 M$_{\odot}$, 1.5 M$_{\odot}$, 2.0 M$_{\odot}$ and 2.5 M$_{\odot}$ stars, starting from the terminal age main sequence to the tip of the RGB. We remove stars in the gray background (\teff $>$ 5500 K) as it is an effective way to sample from the giant branch for $\log g <$ 3.5. Lastly, to make sure we had NUV photometry and parallax distances for our analysis, we only retain stars that were within $3''$ of any entry in the GALEX AIS catalog \citep{Bianchi17} and $2''$ of any entry in the Gaia EDR3 catalog \citep{Gaia21}.The final selected sample consists of 7,286 giants, which we consider throughout the rest of the paper.

\section{Methodology}\label{section:Method}
\subsection{Findeisen \& Hillenbrand (2010) NUV Excess}
To start we adopt the expression of NUV excess from \citet{Dixon20}, which is defined as the NUV-J color displacement from the \citet{Findeisen10} inactive field-star locus. The equation for this locus (Eq. \ref{eq:find_locus}) gives the expected color $(NUV - J)_X$ based on the observed $J - K_s$ color. Following this definition NUV excess is directly calculated via subtraction of this expected value from the observed $NUV - J$. From this point forward we use $E_0(NUV)$ as a shorthand for this definition of NUV excess, as expressed in Eq. \ref{eq:nuv_excess}.

\begin{equation}
    (NUV - J)_X = 10.36(J - K_s) + 2.76
    \label{eq:find_locus}
\end{equation}

\begin{equation}
        E_0(NUV) = (NUV -J) - (NUV - J)_X
        \label{eq:nuv_excess}
\end{equation}

\subsection{Giant Categories for Activity Analysis }\label{ssection:giant_categories}
As a framework for our investigation of activity on the giant branch, we define 5 separate categories based on observed properties. In this subsection we explicitly define these categories, which we refer to frequently in later sections.

\subsubsection{Single and Binary Giants}
To assist our investigation into the role of binarity for active giants we first apply \drvmax constraints to generate single and binary subsamples, where \drvmax is the maximum difference in APOGEE-measured radial velocities for a system. These constraints are \drvmax $<$ 1 \kms for single stars and 3 \kms $<$ \drvmax $<$ 50 \kms for binary stars. These cutoffs yield 4,697 likely single giants and 541 likely binary giants. These limits follow \citet{Mazzola20}, who using simulated APOGEE observations found a cutoff of \drvmax $\geq$ 3 \kms resulted in a completeness fraction of 0.84 and a false positive fraction of 0.005 for binaries in periods less than 100 days. Additionally, they found a completeness fraction of 0.93 using a cutoff of \drvmax $\geq$ 1 km~s$^{-1}$, indicating that giants with \drvmax $<$ 1 \kms are very likely single stars or binaries with relatively wide orbits. We do not make a single or binary classification for cases of only one radial-velocity measurement or \drvmax between the aforementioned constraints to avoid ambiguity. We sample giants with \drvmax $>$ 50 \kms in the following subsection, where we select for binaries that are very likely synchronized. 

\subsubsection{Synchronized Giants}
Assuming that rapid rotation in giants is primarily set by tidal locking with an orbital companion, we expect such giants to be in very short-period binaries. Such binaries tend toward large \drvmax values, depending on the orientation of the orbit and number of measured radial velocities. Most of our sample giants only have a few radial-velocity measurements in DR17, with many only having 2 measured radial velocities. To estimate a reasonable \drvmax cutoff for selecting rapidly rotating synchronized giants, we repeatably sample mock observations of a boundary case. For this case we assume a circular orbit, edge-on orientation and synchronized rotation and orbital periods. Light-curve analyses of mostly main-sequence Kepler field eclipsing binaries found $>70\%$ of them are synchronized at orbital periods $<$ 10 days \citep{Lurie17}. Tidal theory predicts a rapid decrease in synchronization time with increasing radius \citep{Claret97}. This leads to a reasonable expectation that most giants tidally spun up to \vsini $>$ 10 \kms are synchronized. Setting the stellar rotation at \vsini $=$ 10 \kms, a typical literature value delimiting giant rapid rotation, yields an orbital period of $\sim$20 days for a 4 R$_\odot$ giant. This radius is typical of the lower giant branch where most of our sample resides (see section \ref{section:SED}). To determine a cutoff value we draw 2 radial velocities from this boundary case orbit 10,000 times at random phases. The median \drvmax for this exercise is $\sim$50 km~s$^{-1}$, which can be interpreted as a boundary before most giant binaries can be expected to be rapidly rotating. In our sample we find 138 giants with \drvmax $>$ 50 \kms and of those 117 ($\sim$85\%) have \vsini $>$ 10 km~s$^{-1}$. We choose the 117 aforementioned giants to make up our synchronized giant category. Our approach of using both a \drvmax $>$ 50 \kms and a \vsini $>$ 10 is to ensure the giants in this category have strong evidence of both the rotation and nearby companion indicative of synchronized spin up. In this way we select against the case of \drvmax covering the full amplitude of the radial velocity, i.e. longer period orbits where synchronization and thus rapid rotation is not expected. This also selects against rapidly rotating single stars, which may derive spin from accretion of material on the stellar surface or from retention of angular momentum from the previous main-sequence phase. In the case of a main-sequence star with a radiative envelope, spin is likely still significantly reduced by differential rotation \citep{Tayar18}. We acknowledge the contamination we select against is unlikely to be common and that a number of giants in the binary category are probably also synchronized, but justify our cutoffs in order to produce a synchronized giant category that can be treated as highly reliable.

\subsubsection{Sub-subgiants}\label{ssection:SSG}
\begin{figure}[htbp!]
    \centering
    \includegraphics[width=1.1\linewidth]{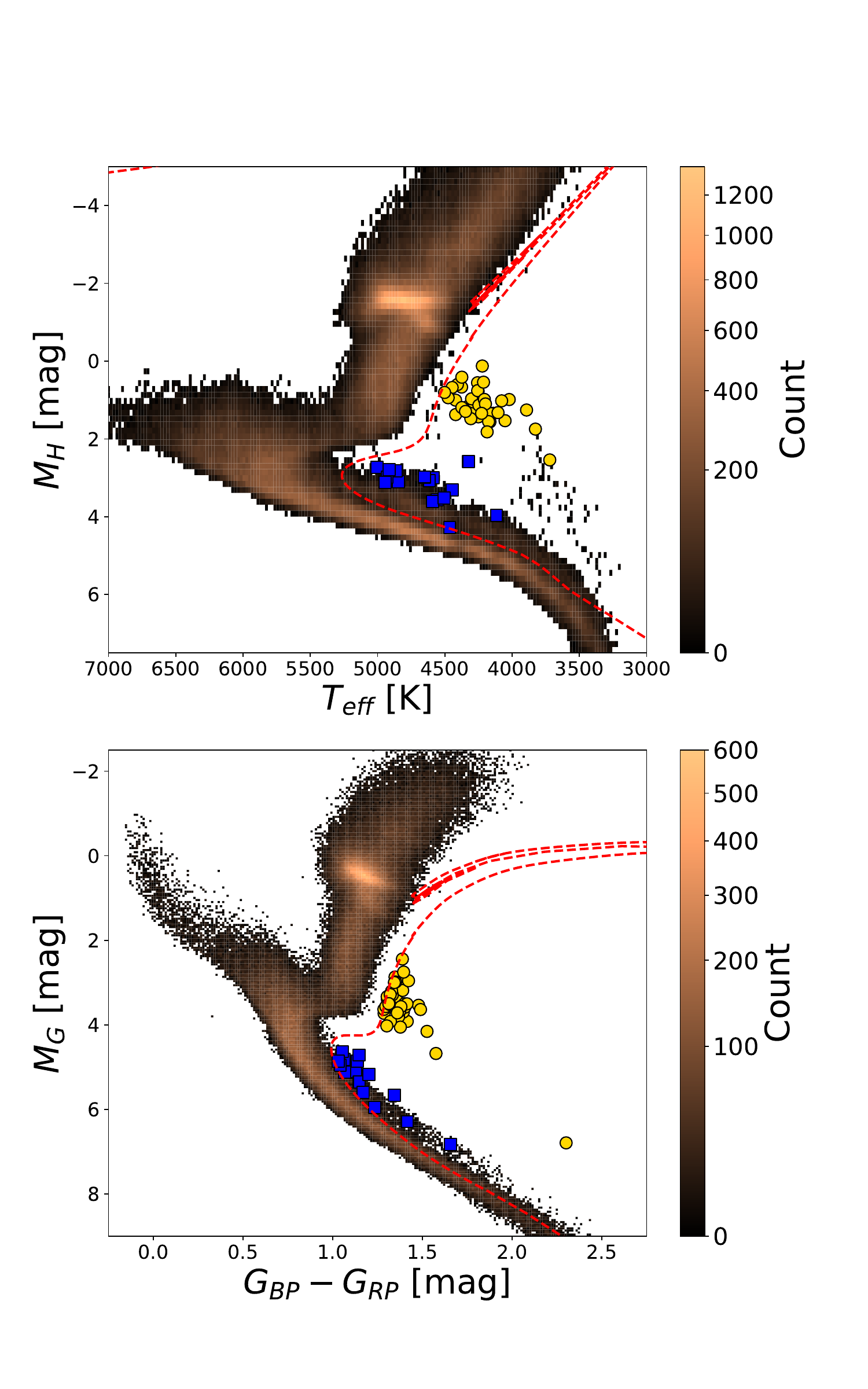}
    \caption{Infrared (top) and optical (bottom) CMD diagrams used to select sub-subgiants. In both panels the histogram is the DR17 catalog and the red dashed line is a 14 Gyr, [M/H] = 0.5 isochrone. The gold circular points are the stars in our sample that are to the red of the isochrone in both CMDs and classified as SSGs. The blue square points are stars redder than the isochrone, but are consistent with the single/binary main sequence so we don't classify them as SSGs.}
    \label{fig:ssg_select}
\end{figure}

Following the approach of \citet{Leiner22}, we used a 14 Gyr [Fe/H] = +0.5 MIST isochrone to conservatively define and select Sub-subgiant (SSG) stars. More specifically, we classified anything to the red of this isochrone in both the infrared and optical as a SSG, excluding points that may be confused with the binary sequence. Given the age of the universe is estimated to be about 13.8 Gyr \citep{Planck2020}, the 14 Gyr isochrone represents a red color boundary for the oldest and most metal-rich giants. Our selection of 38 SSGs can be seen as gold dots for both CMDs shown in Fig. \ref{fig:ssg_select}. The photometry in both panels were corrected using extinction/reddening coefficients from \cite{Zhang23} on E(B-V) values queried from the Pan-STARRS 3D dust map \citep{Green19}. We find our SSGs do not have large extinction values relative to the full sample, and that they reside outside of the Milky Way disk where most star-forming regions exist. This supports that they are unlikely confused with young active stars that can occupy similar regions in CMDs. Interestingly, we find that none of the SSGs are in the single-stars category, but half (19/38) of them overlap with the synchronized category and 14/38 of them overlap the binary category. This is strong supporting evidence that SSGs are generally synchronized short-period giant binaries, which in turn aligns with the hypothesis that SSGs are regularly RS CVn variables, i.e. a class of binary variable stars that typically contains a heavily spotted giant primary component.

\subsubsection{Break-Up Giants}
Using radii ($R$), masses ($M$) and the gravitational constant ($G$) we are able to convert \vsini into \psini and use \citet{Ceillier17} equation 2 to calculate the minimum rotation period before a star breaks up, which is commonly referred to as the critical rotation period (P$_\textup{crit}$). We copy the exact form of this equation in this paper as Eq. \ref{eq:pcrit}.

\begin{equation}
    P_{crit} = \sqrt \frac{27\pi^2R^3}{2GM}
    \label{eq:pcrit}
\end{equation}

\begin{figure}[htbp!]
    \centering
    \includegraphics[width=\linewidth]{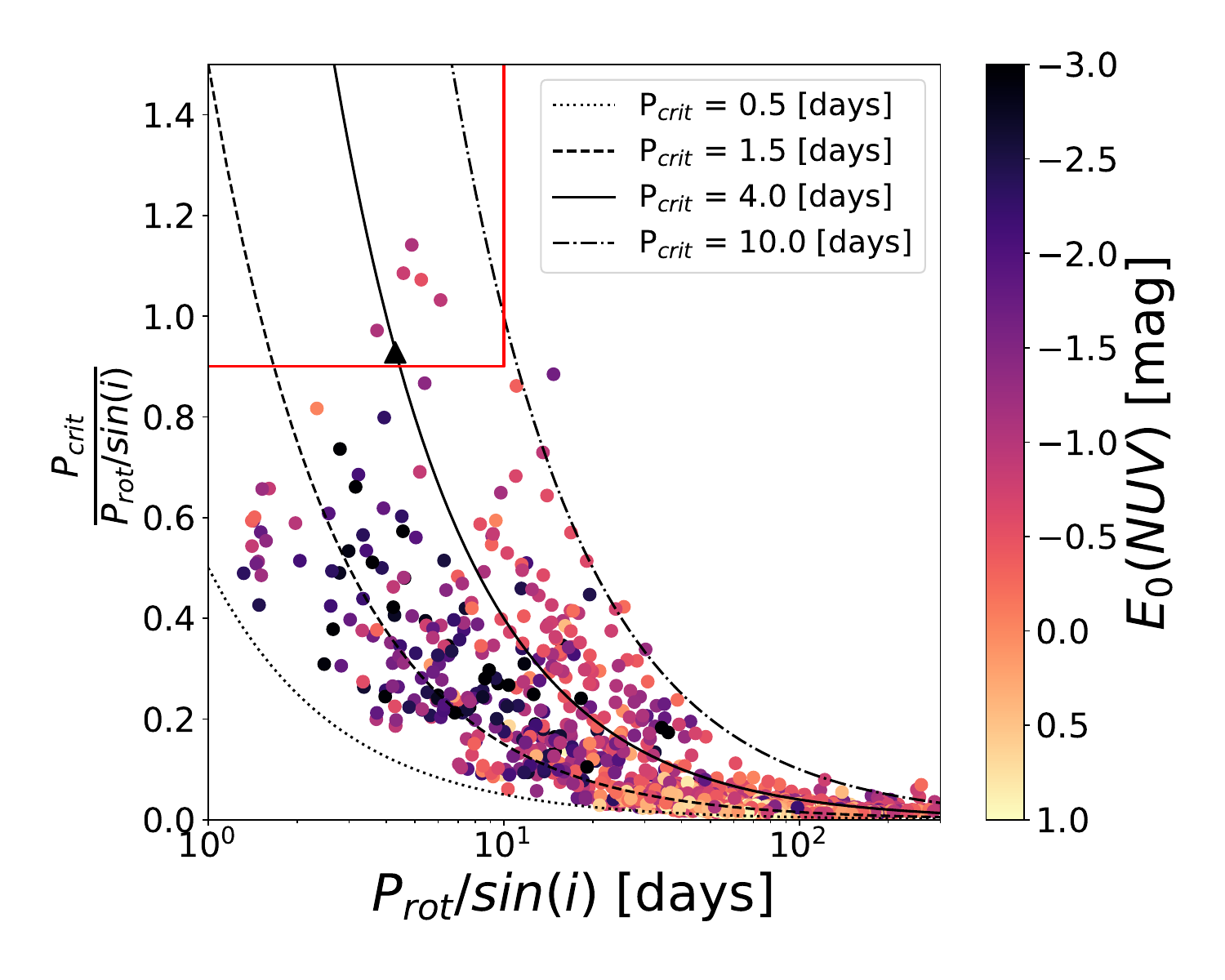}
    \caption{Critical period over projected rotational period versus projected rotational period. To help determine the \pcrit of individual systems in the figure contour lines of constant \pcrit are drawn for reference. The stars in the red box are considered to be critically rotating and categorized as break-up giants. The triangle marker highlights BU giant 2M12072913+0036598, which has $E_0(NUV)$ = -8.45 and is classified as a semi-regular variable.}
    \label{fig:pcrit}
\end{figure}

The radii we use are derived in our SED analysis described in section \ref{section:SED} and the masses are adopted from the APOGEE DR17 DistMass value-added catalog (VAC). This catalog calculates corrected masses from a neural network trained on asteroseismic data \citep{Mosser13} by using ASPCAP T$_{\textup{eff}}$, log g, [M/H], [C/Fe] and [N/Fe] as input vectors \citep{Stone24}.  Fig. \ref{fig:pcrit} shows the ratio of our calculated \pcrit to \psini against \psini. Contour lines are depicted for constant values of \pcrit as a reference for stars sharing similar breakup periods. Filtering only giants with \pcrit values, most (6377/6776) have relatively low fractional break-up periods $\left( \pcritratio < 0.1\right)$. It can be seen that 6 of these giants have outstanding fractional break-up periods that approach or exceed unity $\left( \pcritratio > 0.9\right)$. We categorize them as break-up (BU) giants and attribute apparent supercritical rotation to uncertainties in our radius and mass estimates. We show the propagated uncertainties for \psini and \pcrit with relevant stellar parameters in Table \ref{table:breakup}. Interestingly, the BU giant marked as a triangle (2M12072913+0036598) has a $E_0(NUV)$ of -8.45 and is classified as a semi-regular variable in the AAVSO International Variable Star Index (VSX) catalog \citep{Watson06}. 2M12072913+0036598 was also found to have a P Cygni H$\alpha$ line profile \citep{Tisserand13} and multiple periodic variations in its light curve data. We expound on the properties of these stars during our discussion of potential formation channels for BU giants in section \ref{ssection:BU_Orgin}.

\section{Results}\label{section:Results}
\subsection{NUV Excess \& Metallicity}\label{section:Metallicity}
\subsubsection{Metallicity Effect}
\begin{figure}[htbp!]
    \centering
    \includegraphics[width=\linewidth]{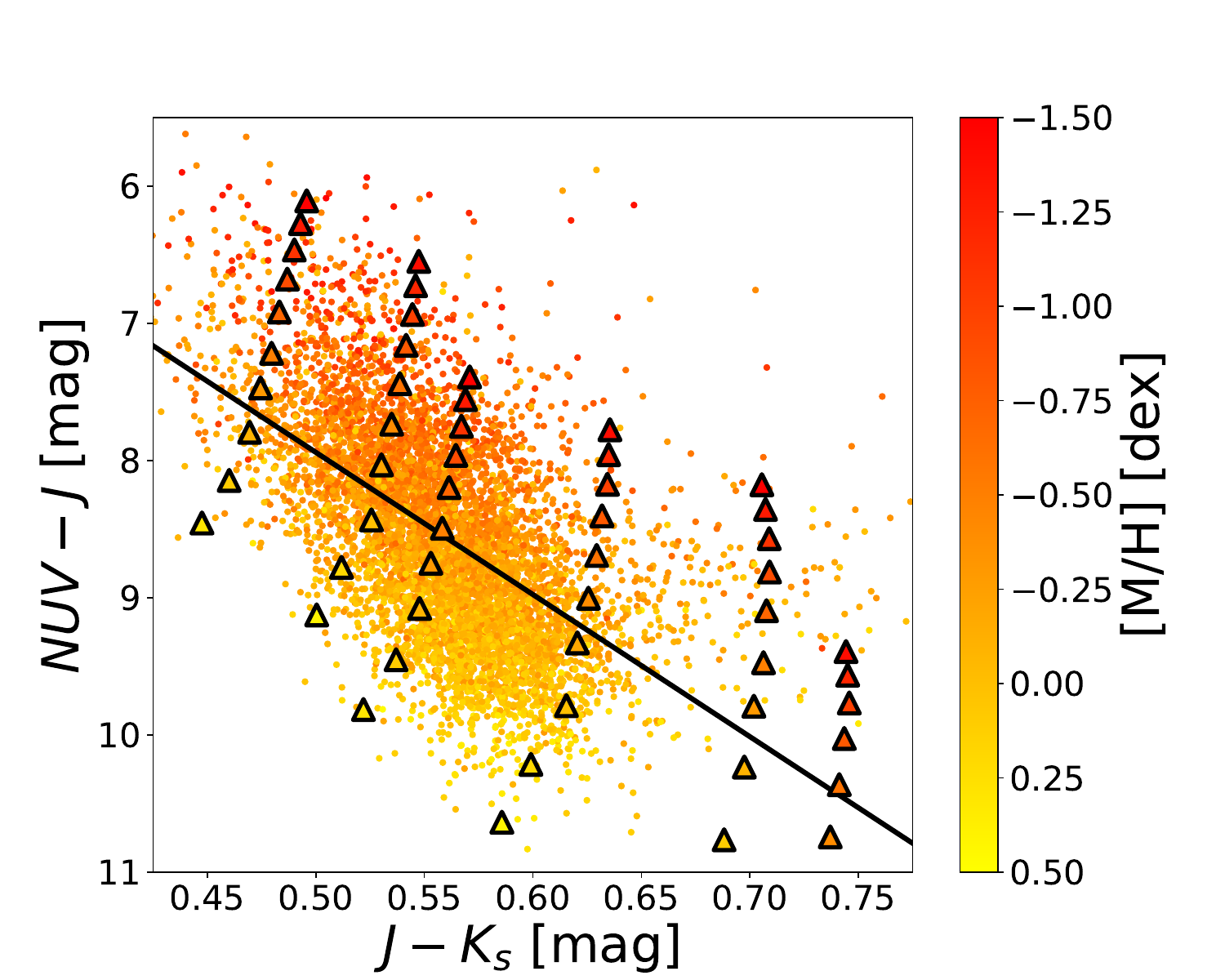}
    \caption{NUV-J versus J-K$_{\textup{s}}$ plot of our sample, colored by [M/H]. A grid of Castelli-Kurucz model atmospheres displayed as triangle markers to show large changes in $NUV - J$ due to} changes in [M/H]. The \citet{Findeisen10} locus is plotted as a black line.
    \label{fig:color_color_ck}
\end{figure}

A statistical comparison between the $E_0(NUV)$ and [M/H] values of our sample reveals the two quantities are correlated to each other with high statistical significance (Pearson r = 0.54, p-value $<$ 1e-6). This correlation has a simple physical explanation as a consequence of line blanketing, where numerous metal lines in the ultraviolet redeposit energy at redder wavelengths \citep{Melbourne60}. We begin our analysis for quantifying this correlation with a comparison of our sample against Castelli-Kurucz (CK) model atmospheres \citep{Castelli03} in $NUV-J$ versus $J-K_s$ color-color space. This is depicted in Fig. \ref{fig:color_color_ck}, where the line represents the \citet{Findeisen10} locus and the overplotted triangles represent synthetic photometric observations of the CK atmospheres. 
The parameter range of the shown models are $4000 <$ \teff $<$ 5100, $-1.5 <$ [M/H] $< 0.4$ and $2 <$ log(g) $< 3$. Changes in model [M/H] cover a range of several magnitudes around the \citet{Findeisen10} locus and mostly wrap around the width of the sample with a similar [M/H] gradient. This is in line with the hypothesis that the correlation between $E_0(NUV)$ and [M/H] can largely be explained by line blanketing.

\subsubsection{SED Analysis and $E_S(NUV)$}\label{section:SED}
To account for changes in $E_0(NUV)$ due to [M/H] we first calculate NUV excess as displaced from a fitted CK model atmosphere rather than the \cite{Findeisen10} locus. The benefit of this approach is that this definition of NUV excess, which we label $E_S(NUV)$, can serve as a more robust activity metric as it is not strongly correlated to [M/H] (Pearson r = -0.11, p-value $<$ 1e-6) like $E_0(NUV)$. This is because the model baseline already accounts for changes in color due to [M/H].  

To perform our SED fitting we use the APOGEE T$_{\textup{eff}}$, log(g), and [M/H] values to generate the model atmospheres and perform a least squares fit of the models to archival broadband photometry in all cases where the group of photometric data points were considered complete. In this context giants were considered to have reliable photometry if none of the magnitudes given for them in the 2MASS and Gaia passbands were null values. None of the GALEX NUV magnitudes could be null either, but that is true for the full sample as a consequence of requiring a GALEX crossmatch in our selection criteria. In total this allows us to calculate $E_S(NUV)$ and stellar radii for 6,640 giants. 

To handle cases with suspiciously low photometric errors, especially in the case of Gaia, we set all errors less than 0.02 mag to 0.02 mag. 

We define our renormalized NUV excess as the $NUV - J$ color displacement from the fitted model. Importantly, the NUV passband is not included during the SED fitting because it is far into the Wien tail for our sample of giants and would mistakenly result in hotter solutions when the dominant source of NUV radiation is from magnetic activity or a hot companion.

We note that the photometry for a significant number of our giants derives from multiple stars, and therefore fitting a single atmosphere is technically invalid. However, due to the large luminosities of the giants we expect the flux contributions of most companions to be negligible in the infrared and optical. A crossmatch to the \citet{Kounkel21} APOGEE double-lined spectroscopic binary (SB2) catalog, which performed a cross correlation search for SB2s in DR17, only returned 4 giants that might be genuine SB2s, although their quality flags indicate potential confusion with spots, reaffirming our expectation of negligible secondary flux contributions. 

\begin{figure}[htbp!]
\centering
\includegraphics[width=\linewidth]{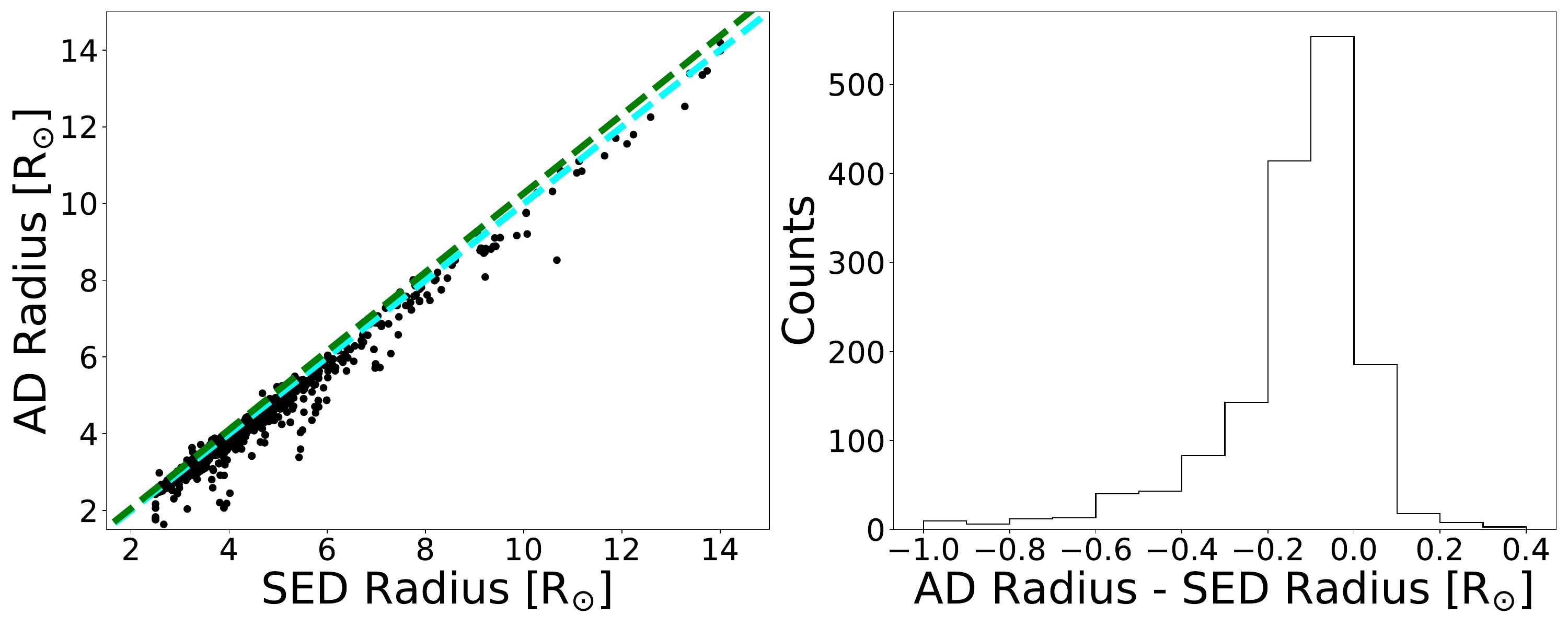}
\caption{Comparison between our SED estimated radii and radii calculated from \citet{Stevens17} color derived angular diameters. A one-to-one line and a robust linear fit line are depicted as the dashed green line and dashed cyan line respectively.}
\label{fig:sed_compare}
\end{figure}

To check the quality of our SED fitting we compare our SED-derived radius and mass values to those estimated using the color-derived angular diameters from \citet{Stevens17}. Out of the 6,640 giants for which we did an SED fit, 1,233 of them also had an angular diameter estimate. We compare the two estimates in Fig. \ref{fig:sed_compare} and highlight where the values are equivalent with a green one-to-one dashed line. We also include a robust linear fit using the Huber loss function to reduce the effect of outliers. This fit is shown in Fig. \ref{fig:sed_compare} as a cyan dashed line. We find an excellent agreement between the two estimates, further validating our SED fitting. We note that there does appear to be a small systematic offset between the radius estimation methods, with a median residual of $\sim 0.10$ R$_\odot$. However, we don't expect this offset to have a significant effect on our results and don't consider it further. This is because the offset is small relative to the calculated giant radii and because the offset is similar in magnitude to the median radius error for both the SED-fitted radii ($\sim$ 0.07 R$_\odot$) and radii determined by angular diameters ($\sim$ 0.11 R$_\odot$).

\subsubsection{$\zeta$([M/H]) Empirical Relation}
$E_S(NUV)$ is an improved activity indicator compared to $E_0(NUV)$ as it does not have strong correlation with [M/H]. However, $E_S(NUV)$ requires SED fitting of model atmospheres and for a given system $E_S(NUV)$ is sensitive to errors introduced from noisy photometry. To deal with these issues and derive a more statistically reliable dependence on [M/H] we fit a linear correction term $\zeta$([M/H]) that can be used to estimate a [M/H] offset for $E_0(NUV)$. This is done by linear regression for $E_0(NUV) - E_S(NUV)$ versus [M/H], excluding giants with NUV excess $< -6$. There are only 5 excluded giants, 3 of which have VSX designations consistent with T Tauri stars (NUV Excess = $-18.9$) and semi-regular variables (NUV Excess = $-8.45$ \& $-12.06$). The final equation for $\zeta$([M/H]) can be seen in Eq. \ref{eq:zeta}.

\begin{figure}[htbp!]
    \centering
    \includegraphics[width=\linewidth]{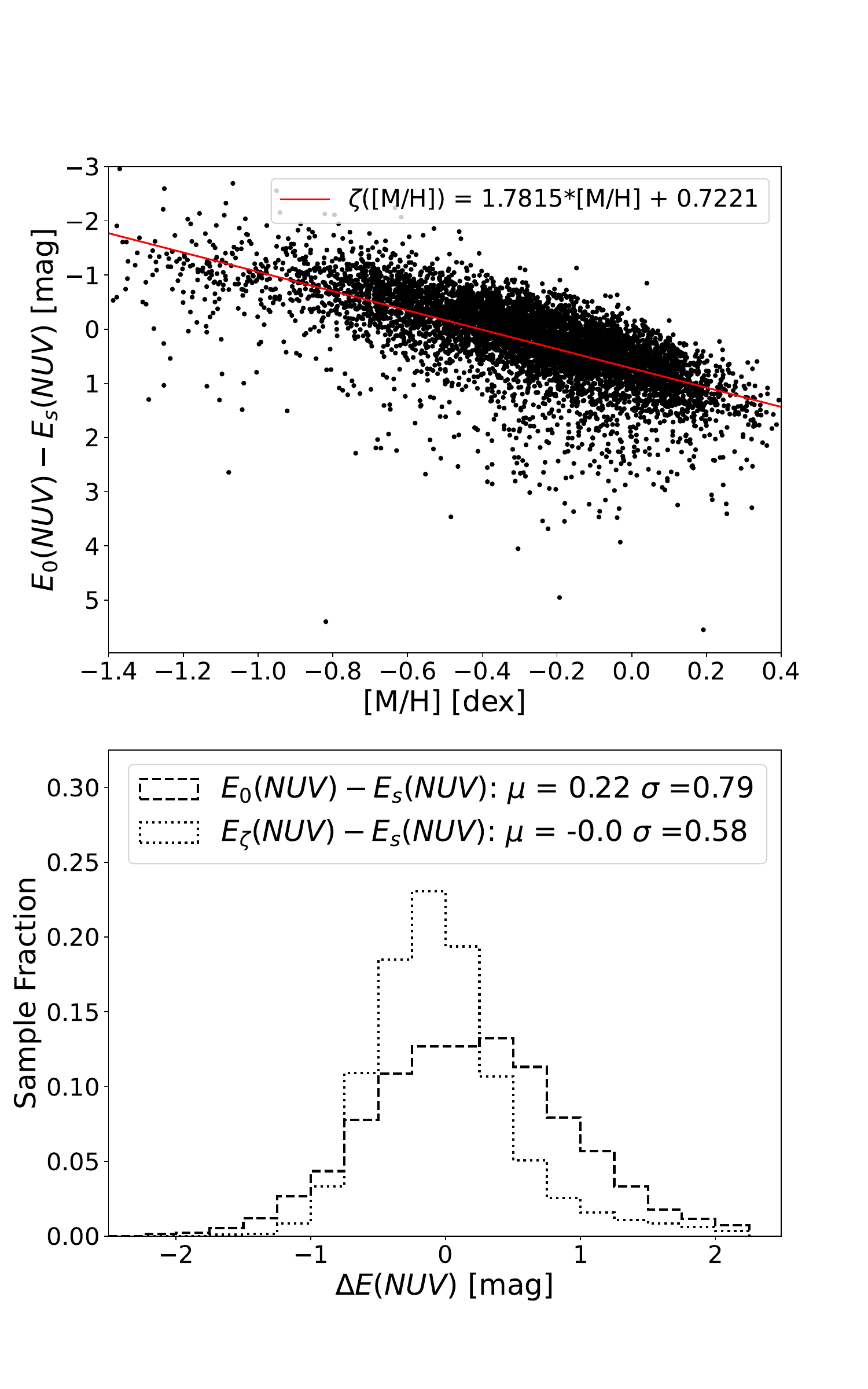}
    \caption{Top panel: Difference between \citet{Dixon20}-defined NUV excess $E_0(NUV)$ and NUV excess defined from fitted CK atmospheres $E_S(NUV)$ versus [M/H]. $E_0(NUV)$ minus the linear regression line $\zeta$([M/H]) yields $E_\zeta(NUV)$ which serves as an estimate for $E_S(NUV)$. Bottom panel: Histograms comparing the $E_0(NUV) - E_S(NUV)$ and $E_\zeta(NUV) - E_S(NUV)$ distributions.}
    \label{fig:norm_nuv_excess}
\end{figure}

The top panel of Fig. \ref{fig:norm_nuv_excess} shows $E_0(NUV) - E_S(NUV)$ versus [M/H] with the best-fit  $\zeta$([M/H]) overplotted. In practice subtracting $\zeta$([M/H]) from $E_0(NUV)$ gives an expected value for $E_S(NUV)$ without the need for SED fitting. This is our $E_\zeta(NUV)$ definition (Eq. \ref{eq:zeta_excess}) and it can be seen in the bottom panel of Fig. \ref{fig:norm_nuv_excess} that it better matches $E_S(NUV)$ than $E_0(NUV)$ alone.

\begin{equation}
    \zeta([M/H]) = 1.7815[M/H] + 0.7221
    \label{eq:zeta}
\end{equation}

\begin{equation}
    E_\zeta(NUV) = E_0(NUV) - \zeta([M/H])
    \label{eq:zeta_excess}
\end{equation}
Lastly, a linear regression between $E_\zeta(NUV)$ and \vsini reveals a relatively good agreement with the \citet{Dixon20} field relation. The results of this regression can be seen in Eq. \ref{eq:zeta_excess_fit}. The slope of the new fit is $-1.200 \pm 0.023$, which is consistent within uncertainties with the slope ($-1.36 \pm 0.16$) of the previous field relation. For all following analysis we use $E_\zeta(NUV)$ as our NUV excess activity proxy. 

\begin{equation}
        E_\zeta(NUV) = (-1.200 \pm 0.023)\vsini +(0.157 \pm 0.011)
        \label{eq:zeta_excess_fit}
\end{equation}

\subsection{The Role of Binaries and Tidal Synchronization}\label{ssection:Orbital_Solutions}
For a comprehensive look into how $E_\zeta(NUV)$ relates to orbital properties of our giants we crossmatched to the Joker value-added catalog (VAC) \citep{Price17} and the two-body non-single star (TB-NSS) table in Gaia DR3 \citep{Gaia23} for values of orbital elements. The Joker is a Monte Carlo sampler designed to find single-lined orbital solutions with a relatively sparse number of radial-velocity measurements. The Joker VAC reports posterior samplings for all APOGEE DR17 giants with three or more visits and that pass system quality checks. In our case we only considered giants with unimodal solutions, i.e. when the sampled orbital elements converge about a single solution. For Gaia we consider both the single-line spectroscopic binary (SB1) solutions and the astrometric (AM) solutions, and removed any SB1 with nonzero quality flags. 

\begin{figure*}[htbp!]
    \centering
    \includegraphics[width=.85\paperwidth]{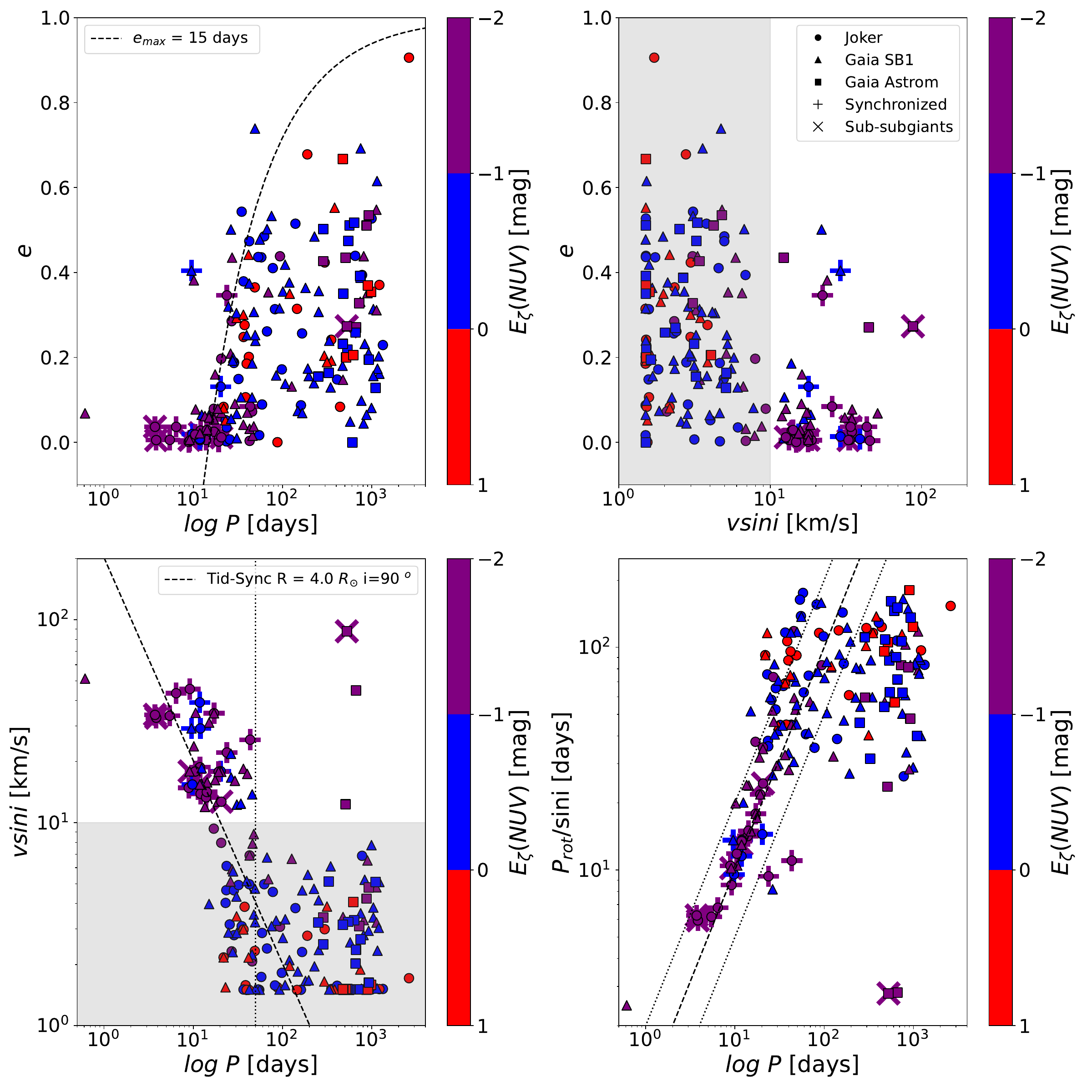}
    \caption{Comparison of ASPCAP \vsini and literature orbital period and eccentricity values. Circles are SB1 solutions from the Joker VAC and triangles and squares are SB1 solutions and astrometric solutions from Gaia DR3, respectively. The X markers are sub-subgiants. All markers are colored by $E_\zeta(NUV)$ divided into three colors with breaks at 0 and -1. The gray backgrounds highlight regions where \vsini $<$ 10 km~s$^{-1}$. In the top left panel a maximum eccentricity curve using a period of 15 days is shown for reference. A key is given in the top right panel to help identify different sources for orbit solutions and giants in our defined synchronized and Sub-subgiant categories. The dashed line in the bottom left panel is a synchronization line assuming a radius of 4 R$_\odot$ and inclination of 90$^o$. The vertical dotted line also in the bottom left panel marks an orbital period of 50 days, which empirically bounds the synchronized giants with \vsini $>$ 10 km~s$^{-1}$. Lastly, in the bottom right panel the dashed line is a one-to-one line with dotted lines shown for factors of 2.  }
    \label{fig:joker_vac}
\end{figure*}

We find 73 Joker SB1 orbit solutions, 173 Gaia SB1 solutions and 31 AM solutions. These giants are represented as circles, triangles and squares respectively in Fig. \ref{fig:joker_vac}. The top left panel compares the orbital period to the eccentricity. Most of the giants with significant eccentricity have orbital periods $>$ 15 days. Initially we discovered a number of very-short-period high-eccentricity Gaia SB1 orbits where this is not the case, but we anticipate that most of these orbit solutions are false. \citet{Bashi22} found such orbits to occur frequently in Gaia SB1 solutions and fit logistic regression coefficients for a handful of Gaia table parameters to calculate validation scores. For giants with  \vsini $<$ 10 \kms we adopt the validation score cutoff of $>$ 0.587 from \citet{Bashi22}, which they found yields a true positive rate of 0.8. Our decision to not remove Gaia SB1 orbits with \vsini $>$ 10 \kms does result in keeping a few likely spurious high eccentricity orbits, which can be seen in the top panels of Fig. \ref{fig:joker_vac}. However, it also preserves a significant amount of apparently synchronized orbits, including some giants we previously categorized as synchronized (+ markers) and/or as SSGs (x markers). 

The top right panel of Fig. \ref{fig:joker_vac} compares eccentricity to $v \sin{i}$, where the gray background highlights stars rotating $<$ 10 km~s$^{-1}$. These ``slow rotators'' have a wide range of eccentricities and often have $E_\zeta(NUV)$ $>$ 0 (i.e., not active) for \vsini $<$ 5 km~s$^{-1}$. Excluding the aforementioned likely spurious Gaia SB1 orbits, there are also a few rapidly rotating giants with eccentric (e $>$ 0.2) astrometric orbit solutions. One of these systems (2M16250995+6645393) overlaps with our SSG selection, has exceptionally high NUV excess ($E_\zeta(NUV)$ $\sim$ -5.17) and is classified as RS CVn variable star in VSX. Unfortunately, none of these three giants have reliable spectroscopic orbit solutions, so we cannot verify the presence of close binary companions. Assuming these giants were tidally spun up, they may currently be in hierarchical triples or perhaps wide binaries where the wide stellar companion dynamically drove the inner binary to merge \citep{Naoz14}. There is also a rapidly rotating giant (2M16032607+0727252) with a significantly eccentric (e $\sim$ 0.35) Joker orbit solution with a period of 23.79 days. Refitting the radial velocity data of 2M16032607+0727252 with our own radial velocity solver we find a somewhat similar maximum likelihood orbit  with a period of 23.81 days and an eccentricity of 0.28. These solutions suggest 2M16032607+0727252 may be surprisingly eccentric compared to other giants with similar rotation, but they are based on only 9 radial velocity visits, and therefore more data is likely needed to finalize the orbital parameters. In general giants with \vsini $>$ 10 \kms are found to be circular or near-circular, and all of them have $E_\zeta(NUV)$ $<$ 0, which does suggest rapid rotation due to tidal synchronization.

The bottom left panel shows that rapid rotation generally occurs in binaries left of the dotted line, which corresponds to orbital periods $<$ 50 days. Excluding the astrometric solution of 2M16250995+6645393, which may be a tertiary of an unresolved inner binary, all of the SSGs strongly follow the trend of synchronization, with \vsini values $>$ 10 km~s$^{-1}$ and orbital periods of $<$ 21 days. They also all have large NUV excess, with $E_\zeta(NUV)$ values $<$ 1. In the bottom right panel we display giants with estimated radii to compare orbital periods to projected rotational period. The dashed line highlights a one-to-one equivalence between both axes, and a factor of two displacement from this locus is depicted with dotted lines. This figure clearly shows strong synchronization at short periods, particularly for giants in our defined synchronized and SSG categories. Curiously a couple of our synchronized giants (2M08195523+2859517 \& 2M16032607+0727252) lie significantly beneath the one-to-one line. Like we did before for 2M16032607+0727252 we refit the radial velocity data, which consists of 10 visits and find the same reported period of 43.30 days.  Additionally, the corresponding best fit SED atmospheres appear to match the data well for both giants. Overall, it appears 2M08195523+2859517 \& 2M16032607+0727252 have orbital periods $<$ 50 days with supersynchronously rotating giant primaries. If these orbital periods survive future scrutiny they might suggest an atypical angular momentum history and investigation into their rotation and orbital properties may serve as interesting tests of binary evolution.

\subsection{Activity on the Lower Red Giant Branch: Rotation, Saturation, and the Extremity of Sub-Subgiants}\label{ssection:Interest}

\begin{figure}[htbp!]
    \centering
    \includegraphics[width=\linewidth]{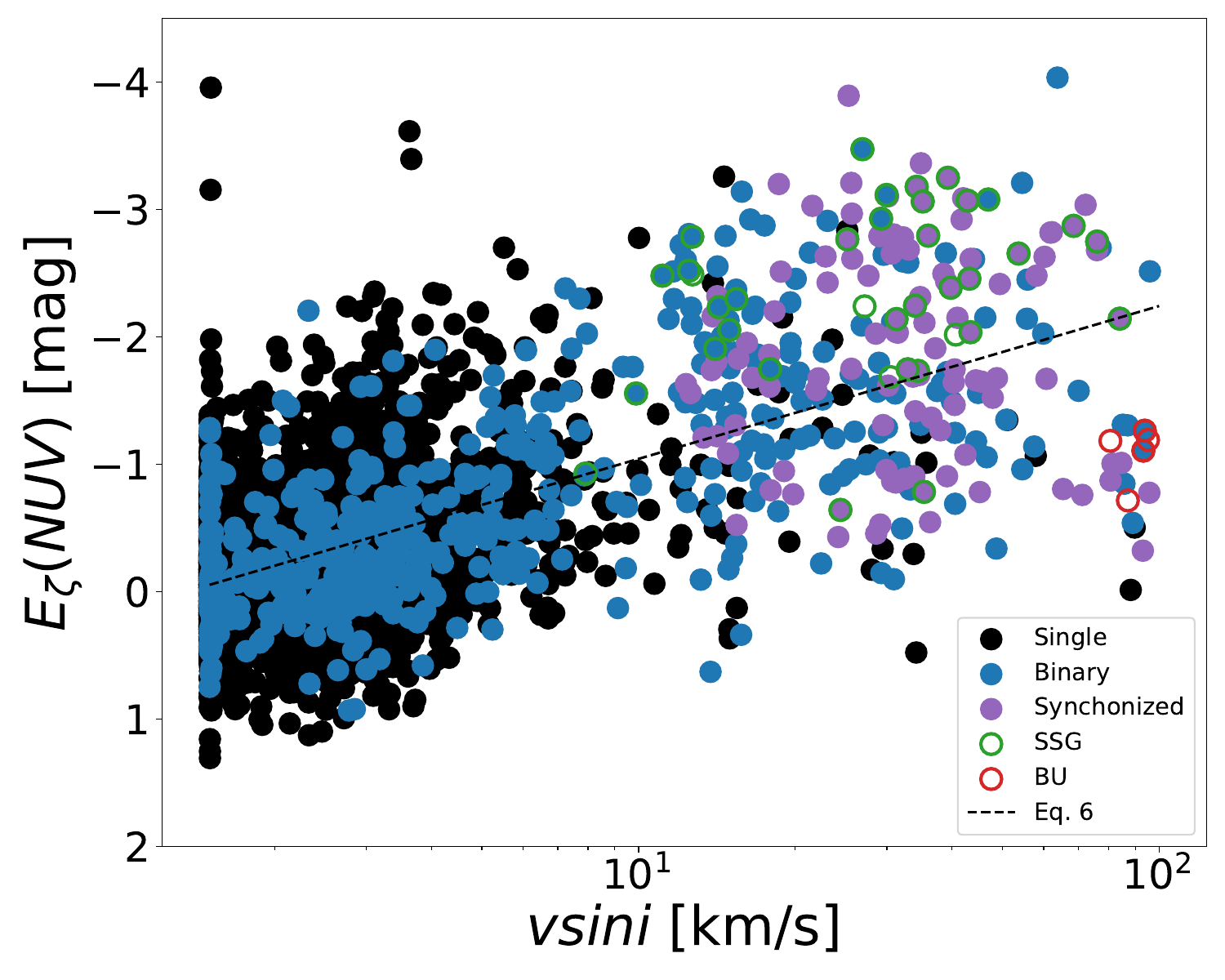}
    \caption{$E_\zeta(NUV)$ vs. \vsini for giant categories defined in section \ref{ssection:giant_categories}. Color scheme for categories is \textcolor{black}{Single}, \textcolor{blue}{Binary}, \textcolor{violet}{Synchronized}, \textcolor{green}{SSG} and \textcolor{red}{BU}. The BU and SSG categories are given as open circle so overlap with other categories can be seen. Dashed line is rotation activity relation as defined in Eq. \ref{eq:zeta_excess_fit}.}
    \label{fig:nuv_excess_vsini}
\end{figure}

Here we provide a narrative overview, based on our findings, of the connection between rotation and activity at the base of the RGB. We start with interpretation of Figure \ref{fig:nuv_excess_vsini}, which is a scatter plot of our $E_\zeta(NUV)$ activity tracer vs \vsini for the categories defined in \ref{ssection:giant_categories}. In the figure single giants are black, binary giants are blue, synchronized giants are purple, and SSGs and BU giants are open green and red circles respectively, so their overlap with other categories can be seen. The giants as a whole scatter around the dashed line, which is the fitted rotation-activity relation (Eq. \ref{eq:zeta_excess_fit}) derived in section \ref{section:SED}. As expected the single stars overwhelmingly ($\sim$ 99\%) have \vsini $<$ 10 km~s$^{-1}$. Interestingly, out of the 45 single giants with \vsini $>$ 10 \kms the 15 with the largest \vsini all lie beneath our rotation activity relation. The largest \vsini for single giants above our activity relation is 25.17 km~s$^{-1}$, while the smallest \psini value is 10.02 days. This result is consistent with the $\sim$ 10 day saturation boundary we found in \citet{Dixon20}. In contrast, the binary and synchronized categories continue to scatter evenly about the relation well beyond \vsini $\sim$ 25 km~s$^{-1}$. This applies to the SSGs especially as 35/38 of them are distributed above our rotation-activity relation. It has been previously observed that some giant binary stars, particularly RS CVn stars, appear “overactive” compared to known rotation-activity relations \citep{Rutten1987}. One explanation for this is that tidal deformations caused by a nearby companion increase the turbulent velocity in tidal bulges, causing additional heating and magnetic activity \citep{Cuntz2000}. Alternatively, it has been hypothesized that spin-orbit resonance can stimulate larger magnetic fields for giant stars \citep{Gehan22}. Whatever the true mechanism is, we suspect the enhanced magnetic activity observed for giant stars in close binaries may explain the lack of saturation we find for them here.

Excluding 2M12072913+0036598 ($E_\zeta(NUV)$ = 8.87) the BU giants are distributed well below our rotation-activity relation and seemingly saturate like the single giants, even though 3/6 overlap with the binary category. Given this and the absence of overlap with the synchronized category despite critical rotation, we suspect the BU giants may not be synchronized binaries at all and likely have comparatively wider orbits.

\begin{figure*}[htbp!]
    \hspace{-35pt}
    \includegraphics[width=.95\paperwidth]{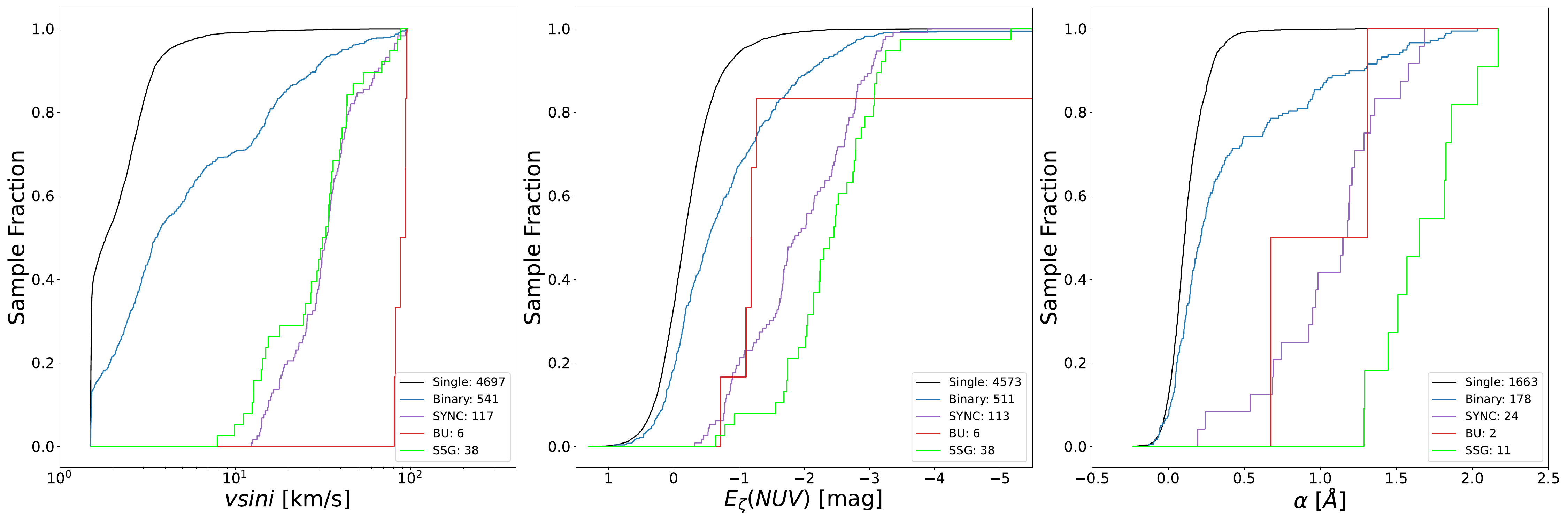}
    \caption{Cumulative distribution functions for \ensuremath{v \sin{i}}, $E_\zeta(NUV)$ and Gaia activity index ($\alpha$) in units of sample fraction for giant categories defined in section \ref{ssection:giant_categories}. Color scheme is identical to Fig. \ref{fig:nuv_excess_vsini}.}
    \label{fig:giants_cdf}
\end{figure*}

Fig. \ref{fig:giants_cdf} shows the cumulative distribution functions (CDFs) of rotation and activity for each giant category. The first two panels show the distributions for $v \sin{i}$ and $E_\zeta(NUV)$ respectively. The third panel shows the Gaia activity index ($\alpha)$ measured from the Gaia Radial Velocity Spectrometer \citep{Lanzafame23}. The index is defined from the excess equivalent width of Ca II IRT lines in a given spectrum. We include $\alpha$ to serve as an independent spectroscopic activity indicator for comparison. The color scheme is identical to Figure \ref{fig:nuv_excess_vsini}.
Beginning with the \vsini distributions, the $\sim$1\% rapid rotation rate we find for single stars is about half of the estimated rapid rotation occurrence rate ($\sim$2.2\%) for RGB stars in the field due to tidal interactions \citep{Carlberg11}. In contrast, the rapid rotation occurrence rate jumps to $\sim$29\% for the binary category, which based on our \drvmax cutoffs mostly consists of orbits $<$ 100 days. Only 2/38 of the SSGs are found to not be rapidly rotating, with the smallest \vsini being $\sim$7.9 km~s$^{-1}$. As expected from the overlap between the synchronized giants and SSGs, their \vsini distributions are fairly similar, with the SSG distribution being a bit more bottom heavy. We suspect at least part of the difference between these distributions are due to our \vsini $>$ 10 \kms constraint of the synchronized category and uncertainties introduced by the distribution of inclinations. The \vsini distribution of the BU stars occupies much faster rotations than our other categories of giants, with \vsini values approaching 100 km~s$^{-1}$.

Examining the $E_\zeta(NUV)$ and Gaia activity index ($\alpha$) panels, which trace activity, we see a similarity in the ordering of the giant categories. The single giants are the least active, followed by the binary category, which tracks with their \vsini distributions. Following this, it can be seen that the synchronized giants are significantly less active than the SSGs in both $E_\zeta(NUV)$ and $\alpha$ despite having an overall slightly faster \vsini distribution. A crossmatch to VSX reveals 12/38 of our SSGs and 24/117 of our synchronized giants have been classified as 
RS CVn stars. If SSGs are generally RS CVn stars, then overactivity and activity cycles between 3 and 20 years \citep{Buccino09,Martinez22} can be expected. 

The $E_\zeta(NUV)$ and $\alpha$ CDFs for the BU giants is also interesting. Excluding 2M12072913+0036598 which resides outside the plot limits, it seems that the BU stars are saturated in both $E_\zeta(NUV)$ and $\alpha$. This suggests further that the BU giants are not synchronized giants and likely gain their angular momentum from means other than tidal interactions. We consider this in detail in the following section.

\subsection{Origin of Break-Up Giants: Planetary Engulfment and Stellar Mergers}\label{ssection:BU_Orgin}
\begin{figure}
    \centering
    \includegraphics[width=.85\linewidth]{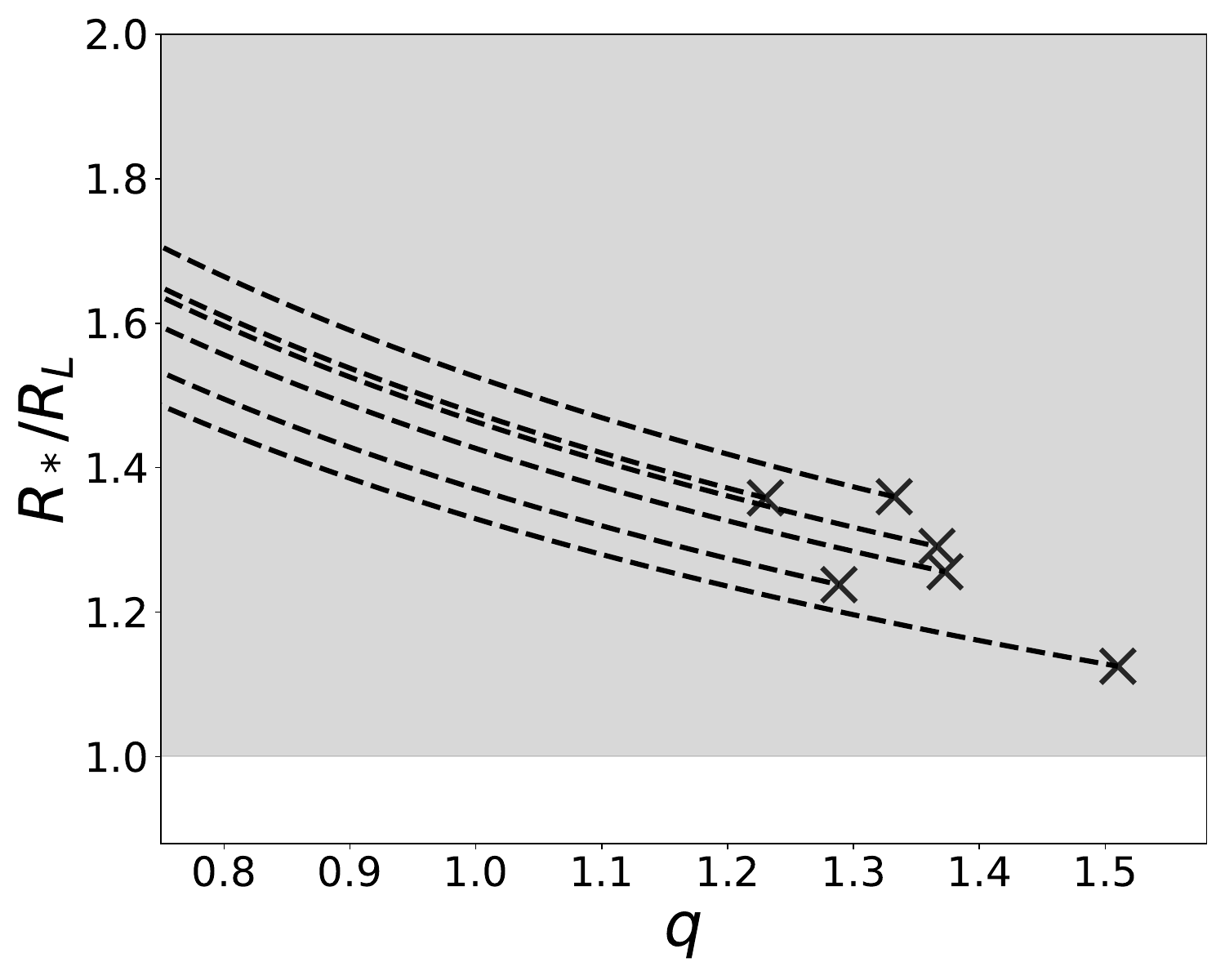}
    \caption{Quotient of stellar radius by Roche limit ($R_*/R_L$) at varying mass ratios ($q$) for BU giants. Gray background highlight region where RLOF occurs. X markers indicate a mass equal to the Chandrasekhar limit for the secondary.}
    \label{fig:rlof}
\end{figure}

For tidal synchronization to spin up RGB stars to break-up rotation, the orbit would generally need to be so close that mass transfer would occur due to Roche-lobe overflow (RLOF). The distance where this occurs depends solely on the stellar mass ratio and orbital semi-major axis \citep{Eggleton83}. If we assume tidal synchronization, then our \psini values are upper limits for orbital periods, and through Kepler's third law this translates to upper limits for orbital semi-major axis. For an array of different mass ratios, we use this value of semi-major axis to calculate how much of the Roche-lobe the six BU giants occupy given their estimated stellar radii and masses (See Table \ref{table:breakup}). This is depicted in Fig. \ref{fig:rlof}, where the gray background above one on the y-axis highlights where RLOF occurs. The x markers truncate the mass ratio arrays where the secondary star is more massive than the Chandrasekhar limit, which shows the range where a white dwarf companion is possible before triggering a Type 1a supernova. This is informative because a more massive companion would have evolved first to become a more luminous giant or a stellar remnant at the time of observation. The case of the more luminous giant is likely unfeasible as it would dominate the light during observations and would thus be the primary, which all stellar parameters are derived for. It would also be larger in size and therefore would induce RLOF even easier. This means that for our case, mass ratio values greater than one would generally indicate stellar remnants, likely white dwarfs.

In summary, Fig. \ref{fig:rlof} shows that even without accounting for tidally distorted surfaces, orbits short enough to synchronize to the BU giant rotation periods would also be within the RLOF domain. This is not necessarily the case for large mass ratios, but then a stellar remnant with a mass greater than the Chandrasekhar limit would be necessary. We conclude that it is very unlikely that the rotation of our BU stars is result of tidal synchronization and that there is probably another mechanism at play.

If the rotation of the BU stars is truly too fast to be caused by tidal synchonization without triggering RLOF, then a reasonable alternative explanation is that their rapid rotation is caused by the angular momentum transport of accreted material. Theory predicts that accreting a small percentage of its total mass is enough to drive a given star to critical rotation \citep{Packet81,Matrozis17,Sun24}. However, mass transfer onto a giant would need to come from a more evolved component, which as previously stated would be the observable component. Additionally, achieving critical rotation in this way for giants is not considered very likely. This is because mass transfer would most likely occur while the accretor was still on the main sequence and therefore envelope expansion and stellar winds would have slowed the rotation to non-critical values.

We consider other mechanisms of accreting matter onto the stellar surface to be more promising explanations. Planetary engulfment for example is another mechanism that can easily spin up a star to critical rotation by depositing orbital angular momentum to the stellar surface, and unlike RLOF mass transfer this is most likely to occur at the base of the RGB where our stars reside \citep{Carlberg09}. A known approach for attempting to identify stars that have accreted planets is the detection of lithium enrichment \citep{Carlberg13,Soares21}. We query against GALAH DR3 \citep{Buder21}, Gaia-ESO DR5 \citep{Hourihane23} and LAMOST DR8 \citep{Zhuohan22}, but we don't find any archival measurements of lithium for our BU stars to check for this signature.

Another avenue for creating critically rotating evolved stars is through stellar mergers. For example, if a binary was close enough the more evolved star could engulf the secondary star as it ascends the RGB, like in the planet engulfment scenario. This could occur as a consequence of the component stars being formed close together or as a consequence of hierachical system dynamics. For example the eccentric Kozai-Lidov mechanism is an efficient three-body interaction for producing very close inner binaries \citep{Naoz14}. This formation channel would also explain why the BU giants seem to be binaries, but do not overlap with the synchronized giants like the SSGs do, despite the extremity of their rotation. This would be the result of observing the naturally wider orbit of the original outer third star in a binary with the merged stellar component.

The merger formation pathway also connects to stellar variability, particularly FK Comae stars. Like our BU giants, \citet{Bopp81} describe FK Comae variables as giants with no close binary companions, extreme rotational broadening (\vsini $\sim$ 100 \kms) and strong magnetic activity. Additionally, \citet{Bopp81} suggest that these stars are a natural consequence of contact binary evolution, where coalescence is predicted to occur during the initial ascent of the RGB. Spectroscopic observations of the prototype FK Comae have revealed double peaked H$\alpha$ emission, which has been characterized by an excretion disk \citep{Ramsey81}. The star has also been found to have non-radial pulsations, thought to be induced as a result of a stellar merger \citep{Welty94}. 

As previously mentioned our BU giant 2M12072913+0036598 is a semi-regular variable, which have been observed to oscillate in non-radial modes \citep{Stello14}. 2M12072913+0036598 is also classified as a cepheid variable in Gaia DR2 photometry \citep{Ripepi19}. We find 2M12072913+0036598 has the largest \drvmax (16.56 {km~s$^{-1}$) and the most radial velocity visits of the BU giants, but suspect the radial velocity variations may be caused by the aforementioned pulsations as our attempts at orbital fitting poorly matched the data. \citet{Tisserand13} performed spectroscopic followup of 2M12072913+0036598 while validating R Coronae Borealis variable star candidates, which are supergiants thought to form from CO and He white dwarf mergers. 2M12072913+0036598 was rejected as a R Coronae Borealis star, but was observed to have a P Cygni H$\alpha$ line profile. This spectral feature and the extraordinarily large $E_\zeta(NUV)$ could be explained with the presence of an excretion disk. These disks have mostly been studied in the context of Be stars, but are known to cause large infrared excess and modeled to have stellar winds powered by rotation and non-radial pulsations \citep{Slettebak88}. The effect of a large infrared excess on $E_0(NUV)$ can be seen in Fig. \ref{fig:color_color_bu_vsx}, where variable stars with circumstellar material are significantly redder than $J-K_s=1.0$ and as a result are extremely displaced in $NUV-J$ over the \citet{Findeisen10} stellar locus. Ultimately we interpret these results as considerable evidence that 2M12072913+0036598 is a product of a recent stellar merger.

 We save a proper evaluation of the relative occurrence of planetary engulfment and stellar merger formation scenarios for critically rotating giants for future work, as we need to perform future observations/analysis to make a convincing discernment between the two formation pathways. 

\begin{figure}[htbp!]
    \centering
    \includegraphics[width=\linewidth]{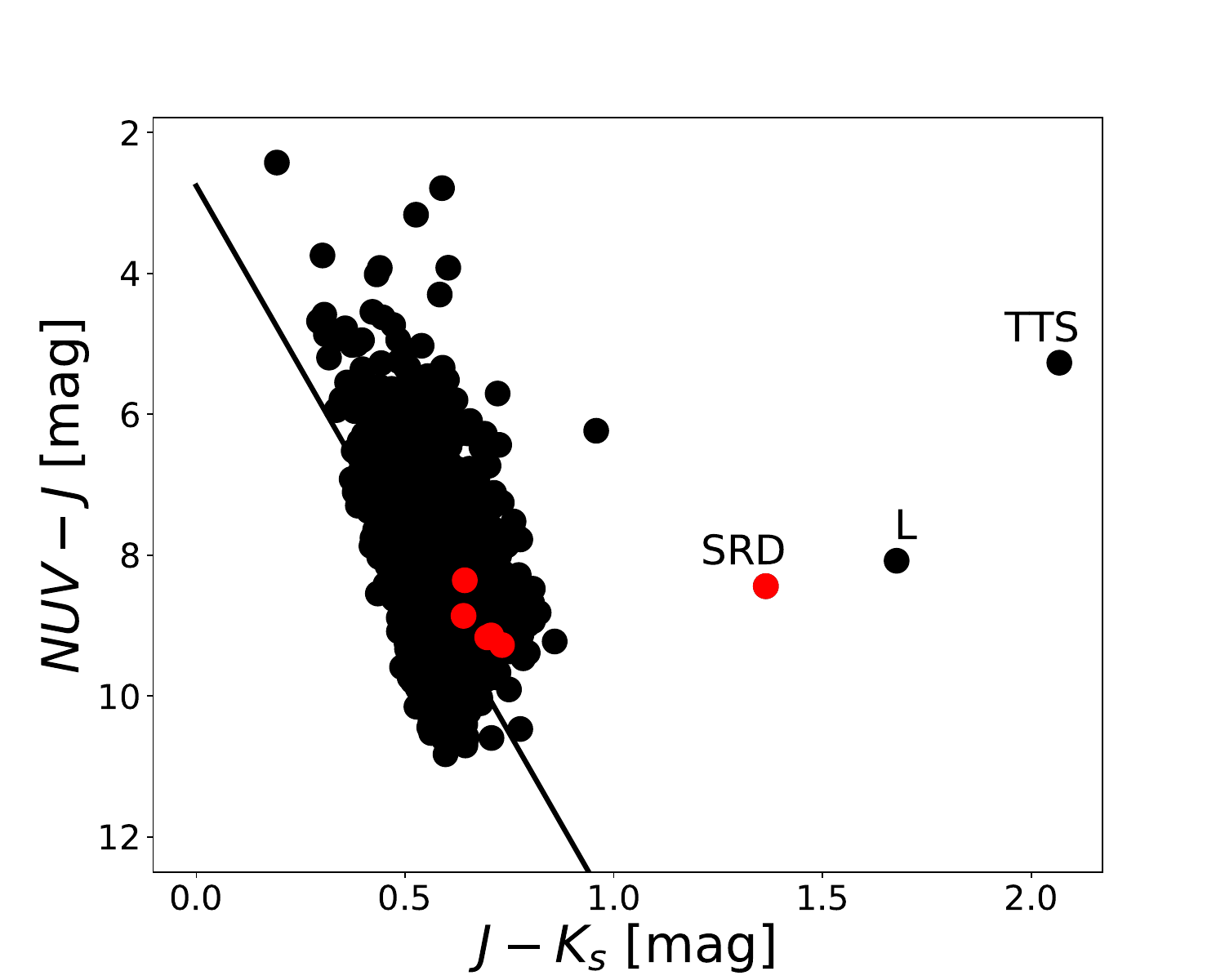}
    \caption{NUV-J versus J-K$_{\textup{s}}$ plot of our sample, with BU giants colored red. The \citet{Findeisen10} locus is plotted as a black line. VSX designations \citep{Watson06} are given for the 3 high infrared excess systems and defined as the following. SRD = semi-regular variable, L = slow irregular variable, TTS = T Tauri star.}
    \label{fig:color_color_bu_vsx}
\end{figure}

\begin{table*}
    \centering
        \def\arraystretch{1.5}
        \hspace*{-3.0cm}\begin{tabular}{|c|c|c|c|c|c|c|c|c|}
        \toprule
        {} APOGEE ID & $T_\mathrm{eff}$ [K] & Radius [R$_\odot$] & Mass [M$_\odot$] & \vsini [km/s]  & \psini [days] & \pcrit [days]\\
        \hline
        2M12072913+0036598 & 4591 $\pm$ 7 & 6.9 $\pm$ 2.4 & 1.0 $\pm$ 0.1 & 81.8 & 4.3 $\pm$ 1.5 & 4.0 $\pm$ 2.1\\
        2M12591606+4206229 & 4679 $\pm$ 8 & 6.9 $\pm$ 0.1 & 1.1 $\pm$ 0.1 & 93.3 & 3.7 $\pm$ 0.1 & 3.6 $\pm$ 0.2\\
        2M16110701+2937086 & 4587 $\pm$ 9 & 9.1 $\pm$ 0.2 & 1.1 $\pm$ 0.1 & 93.8 & 4.9 $\pm$ 0.1 & 5.6 $\pm$ 0.4\\
        2M17022590+3629079 & 4666 $\pm$ 8 & 9.0 $\pm$ 0.2 & 1.1 $\pm$ 0.1 & 87.1 & 5.3 $\pm$ 0.1 & 5.6 $\pm$ 0.3\\
        2M19271456+6852326 & 4519 $\pm$ 7 & 8.6 $\pm$ 0.2 & 1.2 $\pm$ 0.1 & 95.2 & 4.6 $\pm$ 0.1 & 5.0 $\pm$ 0.3\\
        2M23295265+1504155 & 4574 $\pm$ 36 & 9.7 $\pm$ 0.2 & 1.0 $\pm$ 0.1 & 80.6 & 6.1 $\pm$ 0.1 & 6.3 $\pm$ 0.4\\
        \hline
        \end{tabular}
    \caption{Stellar parameters for giants rotating near break-up. Uncertainties for \vsini are not reported in the APOGEE DR17 allstar table and are therefore excluded here.}
    \label{table:breakup}
\end{table*}

\section{Discussion}\label{section:Discussion}
\subsection{Possible White Dwarf Companions}\label{ssection:wd_spots}
One possible explanation for large NUV excess is a hot (young) white dwarf companion contributing large amounts of UV flux, but secure identification of them is difficult. White dwarf companion infrared fluxes would be too dim to be detected as SB2s in the APOGEE infrared spectra. They would also be difficult to distinguish from magnetic activity in our SEDs, as our ultraviolet photometry is limited to the two GALEX passbands and only 248/6,640 of our SEDs have a reported magnitude in the FUV filter. 

To get a handle on the degree of NUV excess contamination from a white dwarf, we perform mock observations on synthetic binaries with a white dwarf component and red giant component. For the giant component we use a 5000 K CK model scaled to varying radii. For the white dwarf component we use $\log g$ = 8 model atmospheres \citep{Koester2010} at varying \teff values scaled to 1 $R_{\oplus}$. The results of these models can be seen in the top panel of Fig. \ref{fig:white_dwarf_companion}, where contours of the giant star model radii are plotted in the $E_0(NUV)$ versus white dwarf temperature space. Cooling ages are also shown on the top axis using DA white dwarf models from \citet{Bedard2020}, which are defined by having thick hydrogen-pure atmospheres. The models clearly show that white dwarfs can contribute significant $E_0(NUV)$ depending on age/temperature, but if we assume that the typical white dwarf is older than $\sim$100 Myr ($<$ 18,000 K), then $E_0(NUV)$ would be limited to $>$ -0.5 magnitudes.

Even for white dwarfs that are very hot ($\sim$40,000 K), $E_0(NUV)$ is still reproducible by the most active giants. It may be true that comparison against spectroscopic activity indicators like the Gaia activity index ($\alpha$) that depend on line formation in the chromosphere could help distinguish between activity and white dwarf companions, but it will most often be necessary to confirm/deny the presence of a white dwarf through other means.

\begin{figure}[htbp!]
    \centering
    \includegraphics[width=.85\linewidth]{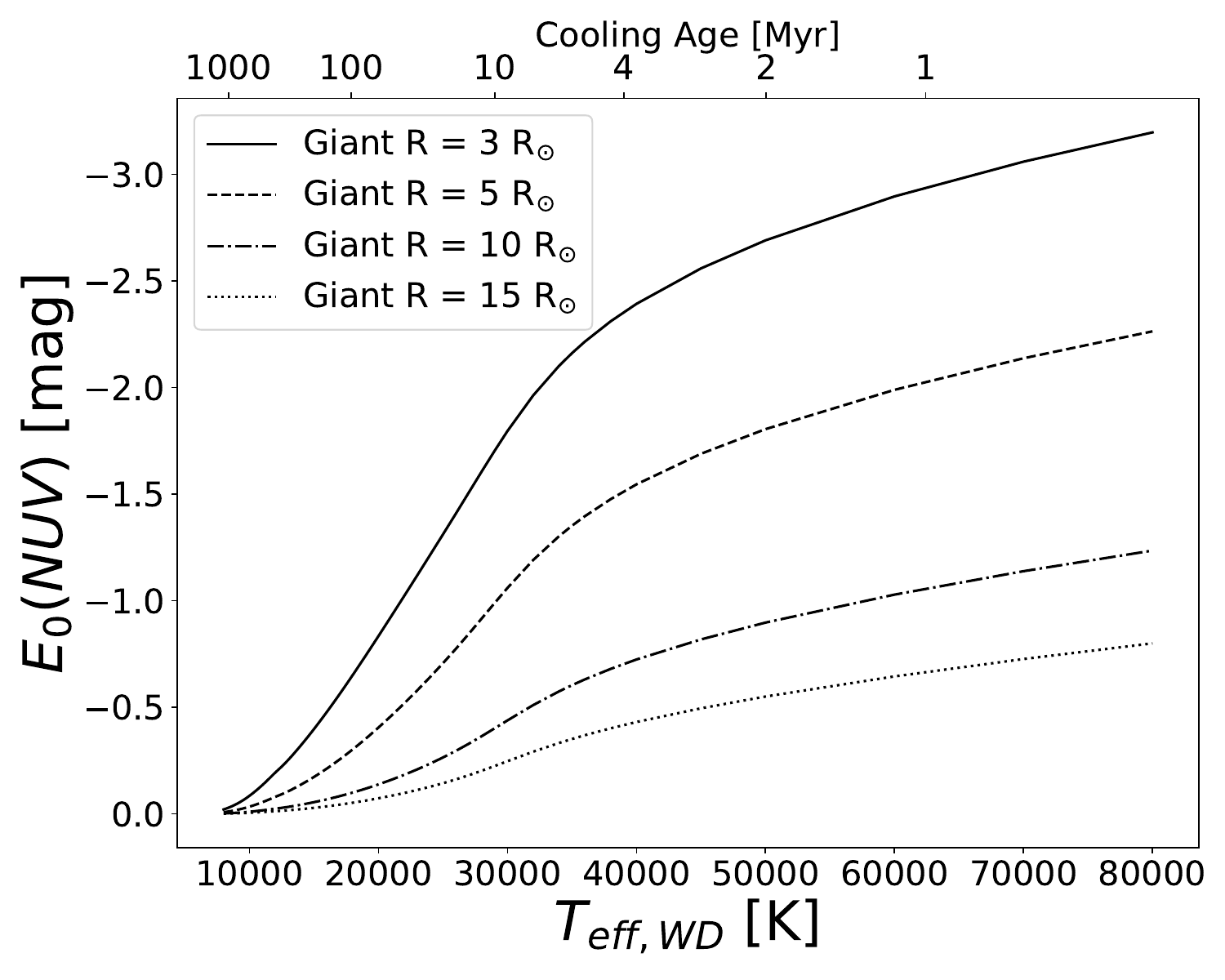}
    \caption{$E_0(NUV)$ vs white dwarf companion \teff for simulated white dwarf giant binaries. Each line represents a 5,000 K giant of varying size and the white dwarf size is set to 1 R$_\oplus$.}
    \label{fig:white_dwarf_companion}
\end{figure}

\subsection{Future Work}
In this paper we have shown that understanding high levels of stellar activity for RGB stars is inextricably linked to rapid rotation and close binary tidal evolution. More specifically, the rapidly rotating UV-bright giants are valuable targets for future work to build our understanding of these connections.

\subsubsection{Light-Curve Characterization}
Rapidly rotating, UV bright giants are particularly ripe for light curve analyses, due to having some combination of the following features imprinted in their light curve data: 1) Heavy spot modulation. 2) Strong ellipsoidal variations. 3) Observable eclipses. 4) Solar-like oscillations. All of these features contain a wealth of useful information about orbital and stellar properties. An example application is comparing the amplitude of light-curve spot modulation against activity proxies like those in this study to garner further insight about the role of activity cycles. Another potential application is further investigation into the role of magnetic activity on modes of stellar pulsations, a topic of recent research \citep{Gaulme14}. 

A quick look at extracted TESS light curves \citep{Oelkers18} for giants with \vsini $>$ 10 \kms and \drvmax $>$ 3 revealed roughly a third (58/179) of them have strongly peaked Lomb-Scargle periods $<$ 13.5 days with normalized power $>$ 0.5. We find many of the light-curve morphologies are consistent with spot modulation, which demonstrates that pronounced rotational variability is very common for short-period active giants. A light-curve analysis to better characterize the entire sample is promising, but we choose to save the completion of this analysis for a future publication in order to be more comprehensive.

\subsubsection{High-Resolution Spectroscopic Followup}
Rapidly rotating, UV-bright giants are also prime targets for future high-resolution spectroscopic follow-up. Although most have only sparsely sampled radial-velocity curves, spin up by synchronization on the lower RGB roughly limits them to orbits ranging from 3.5 days to 50 days. This enables dedicated spectroscopic followup to get good radial-velocity phase converge on times scales of  a few days to several weeks. The SSGs especially are interesting targets as they commonly have the large \vsini necessary to employ Doppler imaging techniques to map the magnetic field and starspot distribution across the surface \citep{Strassmeier04}. In addition to gathering orbital and magnetic field information, SSGs often have short-enough orbital periods for mass transfer scenarios to come into play \citep{Geller17}, possibly resulting in surface pollution detectable through chemical abundances tracers. 

\section{Conclusions}\label{sections:Conclusion}
In this paper we build upon the work of \citet{Dixon20}, which defined empirical rotation-activity relations for 133 APOGEE Red Giant Branch (RGB) stars using \vsini reported from the APOGEE ASPCAP pipeline and NUV excess defined as displacement from the \citet{Findeisen10} stellar locus. Our updated study sample is markedly larger, consisting of 7,286 APOGEE/GALEX giants with ASPCAP \vsini $>$ 0 km~s$^{-1}$, \teff $<$ 5500 K and log g $<$ 3.5. In our analysis we perform SED fitting of Castelli-Kurucz (CK) model atmospheres to this sample and calculate the NUV excess of a respective giant relative to it's best fit CK model. We find the difference between the stellar locus NUV Excess and the CK atmosphere NUV Excess is linearly dependent on [M/H]. By least squares fitting this dependency we derive the $\zeta$([M/H]) equation, which estimates expected NUV excess due to [M/H] alone. In turn, we define $E_\zeta(NUV)$ as $E_0(NUV) - \zeta$([M/H]), which serves as a statistically robust and easy-to-use metallicity corrected UV activity metric.

With this new metric, we again establish a more precise rotation-activity relation for RGB stars that is within errors of the \citep{Dixon20} relation, and thus qualitatively similar to the rotation-activity relation observed among low-mass dwarf stars.

Next, we use the maximum difference in APOGEE-measured radial velocities (\drvmax) to distinguish between giants that are likely to be single or wide binaries and binaries likely to have orbital periods $<$ 100 days. The stars we identify as likely single giants are found to have a rapid rotation (\vsini $>$ 10\kms) occurrence rate of $\sim$1\%. This occurrence rate is about half of the value previously measured for RGB stars in the field, highlighting the essential role of binaries in understanding sources of their occasional enhanced rotation. Our data also shows a saturation regime for $E_\zeta(NUV)$ at \psini $<$ 10 days. This is consistent with the result found in \citep{Dixon20}, except here we also show this applies only to single stars, whereas binary stars show no sign of a saturation regime, indicating binaries can carry substantially stronger magnetic fields.

We find that the sub-subgiants (SSGs) in our sample are generally rapidly rotating, with 36/38 of them having \vsini $>$ 10 km~s$^{-1}$. The SSGs as also mostly in close binaries with 33/38 having \drvmax $>$ 3 \kms and the remaining 5 having no measurable \drvmax due to having only 1 visit. Examination of SSGs with reliable orbit solutions reveal them to generally be synchronized binaries with periods $<$ 21 days. Our findings also demonstrate their activity is measured to be exceptionally high in both $E_\zeta(NUV)$ and Gaia activity index (1.2 $\textup{\AA}$ $<$ $\alpha <$ 2.2 $\textup{\AA}$), even when compared to a more general population of synchronized giants. Additionally, 12/38 of the SSGs are cataloged in VSX as RS CVn stars, which is consistent with the connection made between the two in \citet{Leiner22}. We conclude that SSGs are generally overactive short-period synchronized RGB binaries, which is a description that is almost synonomous with RS CVn stars. The production of large starspots that are typical of RS CVn stars explains the SSGs location in CMDs, as hypothesized in \citet{Leiner22}.

Our analysis also includes 6 giants identified to be approaching critical rotation. Under the assumption of typical mass ratios we show these break-up (BU) giants cannot achieve their rotation rates through tidal synchronization without triggering Roche-lobe overflow. Instead we determine the more likely formation paths for these critically rotating RGB stars are planetary engulfment or stellar mergers. This idea is also supported by the lack of large radial-velocity variations, which would be expected if rotation is driven by tidal synchronization. We note the description of critically rotating single giant stars is very much aligned with FK Comae variable stars, which are thought to be a result of the coalescence of two binary stars during first ascent of the RGB. 2M12072913+0036598 specifically has evidence of nonradial stellar pulsations and an excretion disk, much like the prototypical FK Comae star itself. It is still uncertain to what degree 
the total population of critically rotating giants derive their rotation from mergers or accretion of planets, and so further characterization of these kinds of stars is a good avenue for future research.

In summary this work provides a simple empirical relationship for calculating expected giant activity given $v \sin{i}$, [M/H] and 2MASS/GALEX photometry. Using this equation we show RGB activity is inextricably linked to rotation and binary evolution, primarily though the process of tidal synchronization, but also potentially through planetary engulfment or stellar mergers in the critical rotation case. 

\bibliographystyle{aasjournal}
\bibliography{main}

\end{document}